\documentclass[aps,prl,twocolumn,showpacs,superscriptaddress]{revtex4-1}
\usepackage[latin1, utf8]{inputenc}
\usepackage[english]{babel}
\usepackage{bm,graphicx,graphics,amsmath,amssymb,epsfig}
\usepackage{txfonts}
\usepackage{rotating}
\usepackage[colorlinks,citecolor=blue,linkcolor=Fuchsia]{hyperref}
\usepackage[usenames,dvipsnames,svgnames]{xcolor}
\usepackage{commath}
\usepackage{breqn}
\usepackage{amsmath}
\makeatletter
\let\cat@comma@active\@empty
\makeatother

\begin{document}

\title{Engineering the molecular structure to optimize the spin Hall signal in organics}

\author{M. R. Mahani}
\affiliation{INSPIRE Group, Johannes Gutenberg University, Staudingerweg 7, D-55128 Mainz, Germany}
\email{rmahani@uni-mainz.de}
\author{U. Chopra}
\affiliation{INSPIRE Group, Johannes Gutenberg University, Staudingerweg 7, D-55128 Mainz, Germany}
\affiliation{Graduate School Material Science in Mainz, Staudingerweg 9, D-55128 Mainz, Germany}
\author{Jairo Sinova}
\affiliation{INSPIRE Group, Johannes Gutenberg University, Staudingerweg 7, D-55128 Mainz, Germany}
\affiliation{Institute of Physics, Academy of Sciences of the Czech Republic, 162 00 Praha, Czech Republic}

\begin{abstract}
In this study, by engineering the molecular structure, we optimize the spin Hall conductivity and the spin Hall angle in organics by more than five and three orders of magnitude, respectively. We identify two important characteristics of organic molecules, namely substitution of heavy elements and the torsion angles between constituent units of the polymer, which have significant effects on the spin Hall signal. 
These characteristics are directly related to the spin-orbit coupling and the energetic disorder, both of which offer a wide scope of chemical tunability in high-mobility polymers. 
We compute the spin Hall characteristics for easily synthesized molecules and identify candidates to exhibit the largest spin Hall signals in organic systems, several orders of magnitude larger than previously observed.
The present study brings organic spintronics, by introducing polymers with much stronger spin Hall signal, closer  to their inorganic counterparts.
\end{abstract}
\maketitle 
%
%
%
%
%
%
%
The field of spintronics, takes advantage of the spin degree of freedom of the carriers as well as their charges \cite{wolf2001}. A central focus in spintronics is spin currents and their control by electrical means, instead of magnetic materials or fields. Spin Hall (SH) effect and its inverse counterpart, the inverse spin Hall effect (ISHE), provide the means to create and detect the spin current and provides a direct measurement of spin information \cite{sinova2015}. In the SH effect, a pure spin current is generated in the transverse direction to the charge current passing across the system, while, a spin current is detected by the conversion to a transverse charge current in the ISHE. The SH effect for crystalline systems with itinerant electrons has been studied extensively \cite{sinova2015, sinova2004}. However, it has not been fully explored for the disordered systems with localized electrons, such as organic materials.

Organic materials have lighter element composition as opposed to conventional semiconductors, which accounts for the weaker spin-orbit coupling (SOC) in the former. Although weaker SOC provides a long spin relaxation time or spin diffusion length, it negatively influences the SH signal. This has contributed to the difficulties in the observation of the SH signal in organic materials. On the other hand, organic materials are low cost, easily synthesised, solution processable, and flexible. This gives the organic spintronic devices unprecedented tunable possibilities. The tunable molecular structure allows us to synthesize materials with optimal specifics to enhance SH effect.

%
%
SH effect was not explored in organics till only a few years ago \citep{ando2013}. Most of the theoretical studies were focused either on extrinsic SH effect or intrinsic effect for materials whose dominant transport mechanism is band transport, namely, crystalline inorganic metals and semiconductors with high mobility \cite{Dyakonov1971, hirsch1999, sinova2004, Entin-Wohlman2005, sinova2015}. The few studies done in organics have proven the observation of the SH effect challenging, due to thermovoltages comparable in symmetry and magnitude to the SH signal. After the first observation of an inverse spin Hall voltage in a solution-processed conducting polymer \citep{ando2013}, 
thermovoltages similar to the SH signal were reported, brining into questions the origin of the previously measured signals
\citep{wid2016, wang2018}.  Very recently, we have shown that, there is indeed a very small but finite inverse SH signal that can be measured in organics, if thorough examination of various artefacts, including thermovoltages, and a specific sample design is considered \cite{mahani2020}. Here, we show molecular-structure engineering which can enhance the very weak SH signal in organics by orders of magnitude.

%
%
%
In this study, we calculate the SH angle on easily synthesizable organic molecules, with parameters extracted from first principles. In the following, we briefly describe the theoretical model and the parameters which influence the SH effect. We show polymer characteristics which can improve the SH angle by more than three orders of magnitude. This brings us closer to a SH effect in organics comparable to that of inorganic semiconductors.

%
%
%
%
%
Our theoretical calculations for the SH conductivity and the SH angle are based on a recent theoretical work on the SH effect \cite{Yu2015c} and the anomalous Hall effect in disordered materials \cite{Liu2011b}. In the hopping regime, the SH effect, similar to the ordinary or the anomalous Hall effects, arises when in addition to the hopping between pairs of sites (site i and j), the hopping in triads (hopping via an intermediate site k) is also considered. In an organic material in which the molecular orientations are not aligned, the hopping integrals over a triad loop for the SH effect, give rise to a non-zero phase shift, due to the presence of SOC \cite{Yu2015c}. The importance of hopping in triads was first recognized by Holstein~\cite{Holstein1961} and was subsequently shown by others~\cite{bottger1976hoppingI, bottger1976hoppingII, friedman1978hall, butcher1980hall, friedman1981hall, butcher1981hall, bottger1982hopping, butcher1983hall} in the study of the ordinary Hall effect and later for the anomalous Hall effect~\cite{Liu2011b, sinova2015}.

In this work we consider a system with more than 10 000 sites, arranged on a cubic mesh. Hopping between any pairs and triads of sites within a cutoff distance is taken into account. With parameters taken from our first principle density functional theory (DFT) calculations or literature values, we obtain the electrical conductivity by applying a small bias in a resistance network, within the linear response regime, across the x or y direction. First, the voltage drop along the x or y direction is obtained using
\begin{equation}
\sum_j{g_{ij}(V^{x/y}_j-V^{x/y}_i)}=0,
\label{vdrop}
\end{equation}
where the conductance between site i and j is defined as $g_{ij} =\nu e^{-2\alpha |R_i-R_j|}e^{-(\beta/2)(|\epsilon_i|+|\epsilon_j|+|\epsilon_j-\epsilon_i|)}$ and $V^{x/y}_i$ are the potentials (See supplementary information). The parameters $\epsilon$, $|R_i-R_j|$, $\alpha$ and $\nu$ are the site energies, the inter-site separation, the decay constant of the localized wave functions and the attempt-to-escape frequency, respectively. The electrical conductivity is calculated then by \cite{butcher1977analytical}: 
\begin{equation}
\sigma_{xx/yy} =\frac{1}{2\Omega E^2} \sum_j{g_{ij}(V^{x/y}_i-V^{x/y}_j)^2},
\label{econd}
\end{equation}
with $\Omega$ and $E$ being the volume of the system and the electric field, respectively.
Extending the theory of the AHE in the hopping regime \cite{Liu2011b} to the SHE \cite{Yu2015c}, the SH conductivity is obtained via $\sigma_{sh}=J_s^y/E^x$, with $J_s^y$ the spin current along y direction when the electric field, $E^x$, is applied along the x direction. The SH conductivity is written as
\begin{equation}
\sigma_{sh} =-\frac{e^2\beta}{6\Omega E^2} \sum_j{W^{z}_{ijk}(V^x_{ij}V^y_{jk}-V^y_{ij}V^x_{jk})},
\label{shcond}
\end{equation}
where the indirect hopping probability via an intermediate site is given by
\begin{multline}
W^{z}_{ijk} =\frac{\lambda N^z_{ijk}}{V_0}\hbar \nu^2 e^{-\alpha(R_{ij}+R_{jk}+R_{ki})}[e^{-(\beta/2)(|\epsilon_j|+|\epsilon_k|+|\epsilon_j-\epsilon_i|+|\epsilon_k-\epsilon_i|)}+\\
i\rightleftharpoons j +i \rightleftharpoons k].
\label{thp}
\end{multline}
The SH angle is determined as $\Theta_{sh}=\sigma_{sh}/(\sigma_{xx}\sigma_{yy})^{1/2}$ with $\sigma_{xx}$ and $\sigma_{yy}$ as the electrical conductivity along the x and y directions, respectively.

In order to ensure that our results are not the artefact of an ordered cubic mesh, we allow for a spatial disorder by altering the distance between any pair of sites randomly in all directions by up to 15$\%$ of its original distance. Within this theoretical framework, we identify the important parameters which have significant effect on the SH conductivity or SH angle. Each parameter is related to a molecular characteristic that can be tuned to improve the SH angle. 

We performed DFT calculations to obtain the SOC parameter  at the TZVP/PBE0 level of theory using NWChem version 6.8 \cite{Valiev2010}. The geometry optimization of the polymers was done in cationic state with a single charge delocalized over two units. Spin-admixture calculations were performed at the same level of theory based on the formalism described in detail in Ref. \citenum{chopra2019accurate}. We calculate the site-energies using DFT for individual segments of the polymers, which were obtained from a snapshot of a molecular dynamics simulation for a semi-crystalline sample. The energetic disorder is then obtained by the standard deviation of the distribution of the site-energies.

%
%
%
%
%
%
%
As we show in the following, the SH angle in organics can be improved by more than three orders of magnitudes, using the molecular structure. We identify that the energetic disorder and the SOC have the strongest influence on the SH conductivity and the SH angle. In order to study the effect of these parameters on the SH effect, we have chosen molecules where we can systematically vary either of these properties without influencing the other. For instance, poly(2,5-bis(3-alkylthiophen-2-yl)thieno(3,2-b)thiophene), PBTTT, and PBTe differ only in elemental composition (replacing Sulfur with Tellurium), thus we expect a larger SOC in PBTe while both having a similar charge transport and energetic disorder (confirmed by our DFT calculations). On the other hand, indacenodithiophene-co-benzothiadiazole (IDTBT) and PBTTT have a similar range of spin-admixtures but their calculated energetic disorder is quite different.\\
\begin{figure}
\centering
\begin{minipage}{0.45\textwidth}
\includegraphics[width=\textwidth]{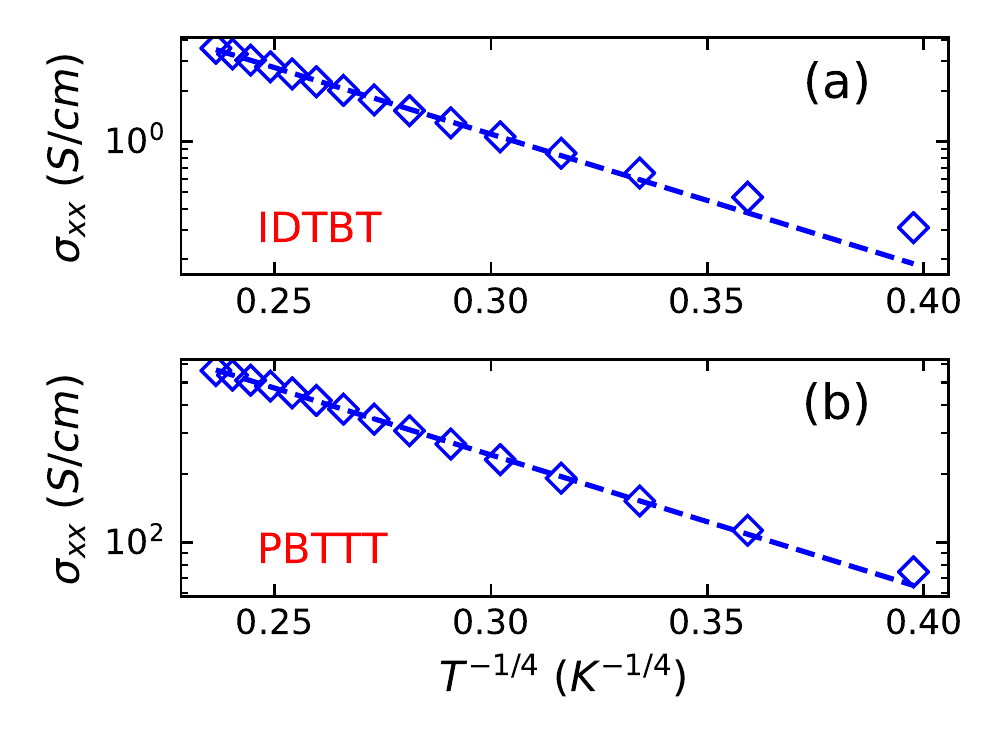}
\end{minipage}\hfill
\hspace{0.50mm}
\begin{minipage}{0.45\textwidth}
\includegraphics[scale=0.15]{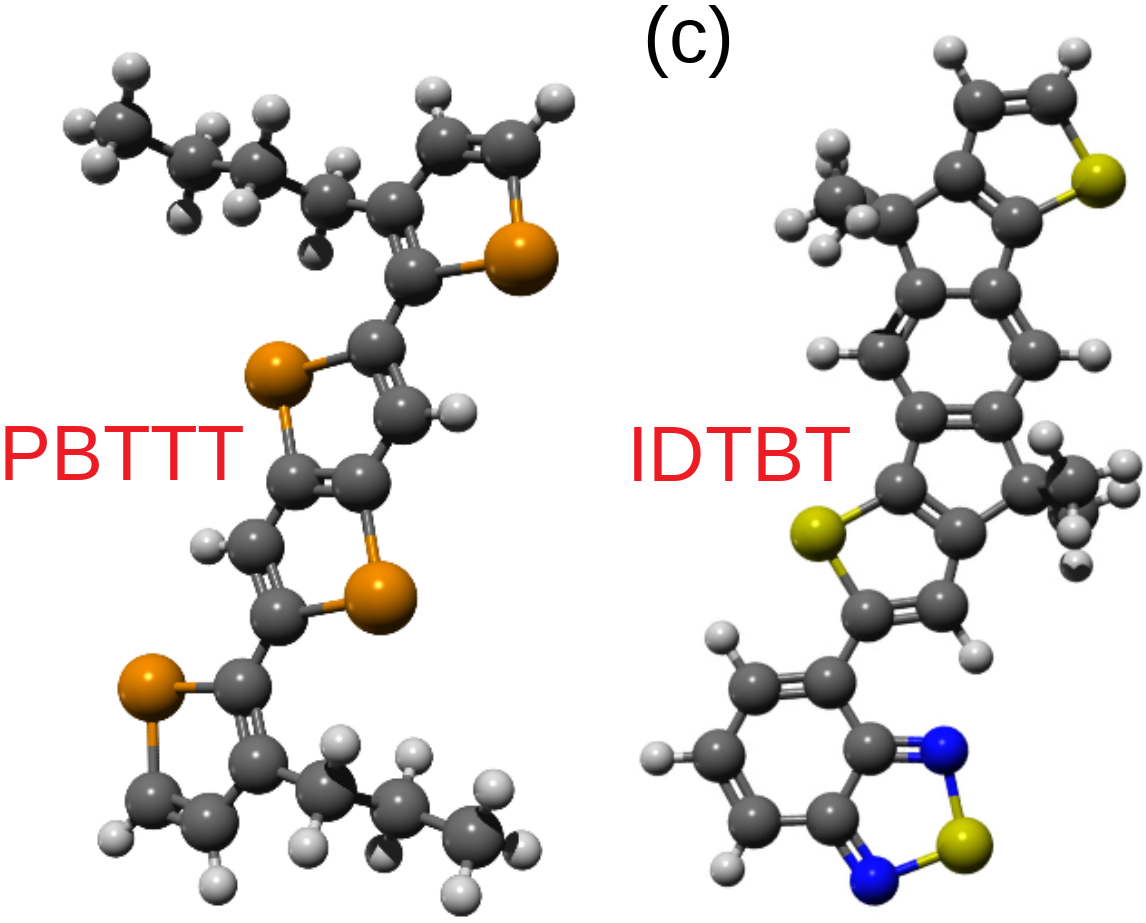}
\end{minipage}\hfill
\caption{Electrical conductivity $\sigma_{xx}$ as a function of temperature for (a) IDTBT and (b) PBTTT. (c)The chemical structure of IDTBT and PBTTT. The line is fitted to the conductivity data-points for 3D variable range hopping transport.}
\label{fig:elec_cond_idtbt}
\end{figure}
\begin{figure*}
\begin{minipage}{1.0\linewidth}
\centering
\includegraphics[scale=0.75]{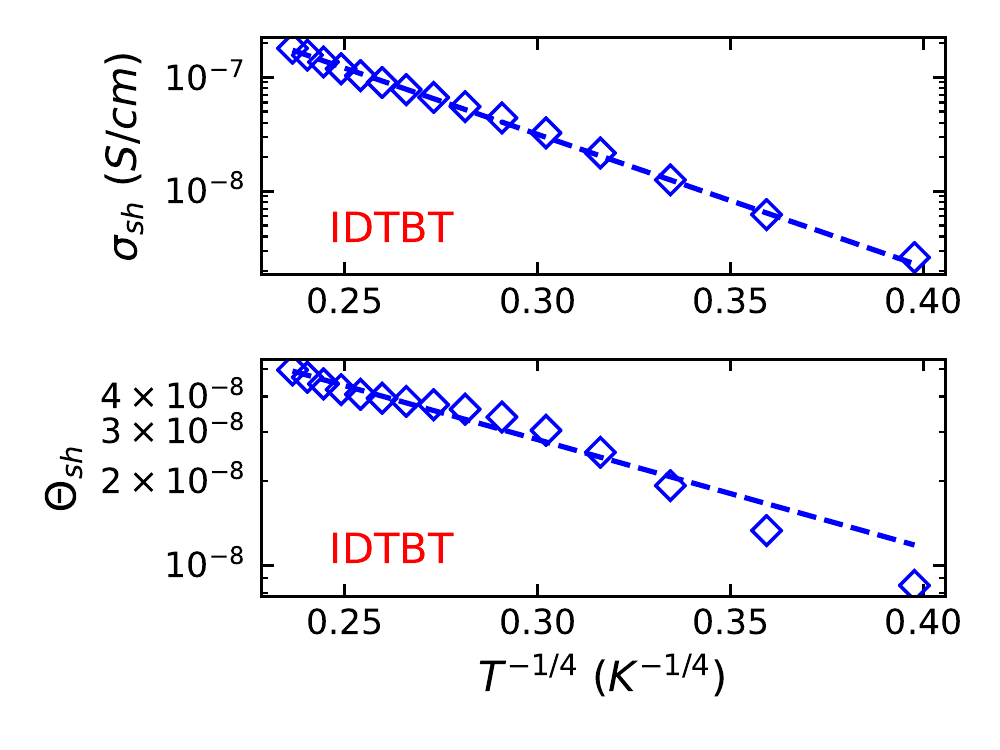}
\hspace{-3.50mm}
\includegraphics[scale=0.75]{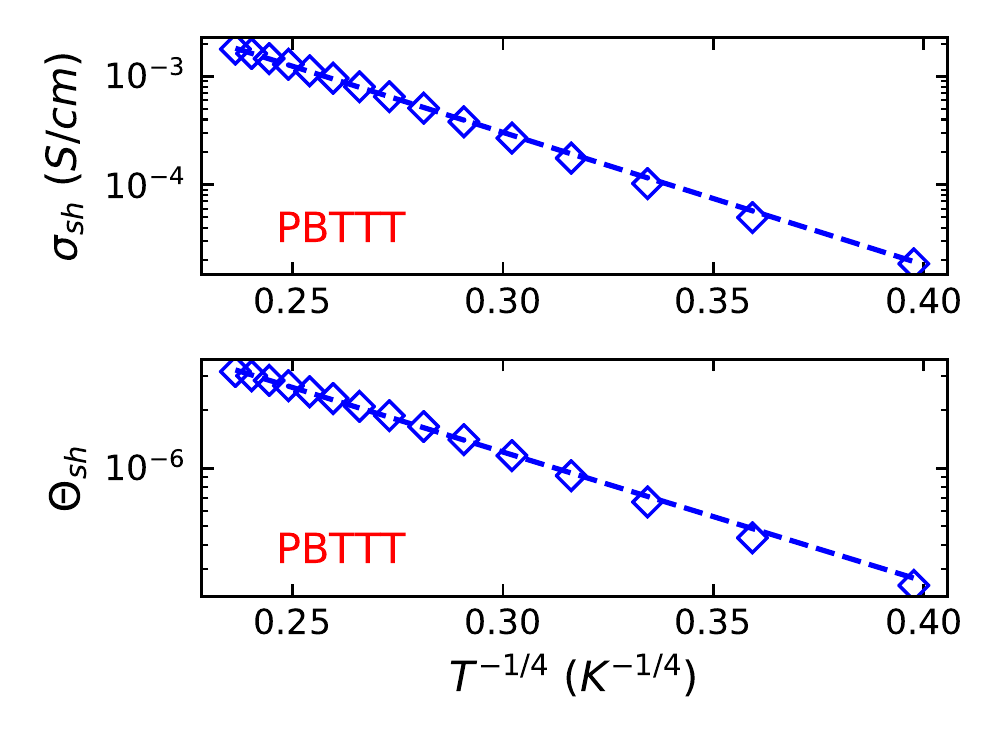}
\end{minipage}
\vspace{0.00mm}
\caption{The SH conductivity $\sigma_{SH}$ and the SH angle $\Theta_{SH}$ as a function of temperature for IDTBT and PBTTT, two polymers with different energetic disorder. The temperature ranges from [40, 320] K.}
\label{fig:ED_comp}
\vspace{-0.5 cm}
\end{figure*}

We start by analyzing the electrical conductivity and its signatures in IDTBT and PBTTT. These two polymers have rather similar structures, similar charge transport properties in the crystalline regions \cite{venkateshvaran2014approaching} and both possess high mobilities. On the contrary, they have different energetic disorders, with PBTTT showing the lowest degree of energetic disorder. PBTTT is a semi-crystalline conjugated polymer in which the lamellar micro-structure is highly ordered. The high in-plane mobility in this polymer stems from the fact that the plane in which the $\pi-\pi$ stacking direction and polymer chains are oriented, is roughly perpendicular to the alphabetic side chains. The experimental data on the electrical conductivity in PBTTT differs due to different characteristics, while there are not many experimental data on IDTBT. For IDTBT, a 3D variable range hopping (VRH) transport with maximum electrical conductivity of 8 S/cm at room temperature has been reported \cite{bei2015electronic}. For PBTTT, a 3D-VRH transport among the localized states has been reported \cite{ito2016critical}. The same group has also reported a 2D-VRH transport with the maximum conductivity of 300 S/cm at the doping concentration of $10^{21}$ cm$^{-3}$ \cite{harada2015signature}. The difference was attributed to the different thicknesses used in two experiments \cite{ito2016critical}. In a high-performing PBTTT thin film, a high electrical conductivity of 670 S/cm has been reported \cite{patel2017morphology}.

Our calculations for the electrical conductivity of IDTBT and PBTT as a function of temperature is plotted in Fig.~\ref{fig:elec_cond_idtbt} (a) and (b). The chemical structure of the molecule is shown in Fig.~\ref{fig:elec_cond_idtbt} (c). Since our system is identical along the x and y direction, the electrical conductivity is also alike and therefore, we only plotted the conductivity along the x direction. In this figure, the blue diamond data-points are the calculated electrical conductivity and the lines are fitted according to 3D-VRH transport. Both fitted lines mach the data points perfectly from temperatures above 80 K up to room temperature, while for the lower temperatures, the fit is reasonably good. 
The maximum electrical conductivity at room temperature for both PBTTT and IDTBT is within the range of experimental values reported above.

Figure \ref{fig:ED_comp} shows the effect of energetic disorder on the SH signal. Energetic disorder relates to the distribution of the site energies around the Fermi energy, indicated by $\epsilon$ in Eq.~\ref{thp}. The site energies are distributed randomly within an interval $[-\epsilon/2, \epsilon/2]$, showing that the narrower the interval in energy, the stronger the SH conductivity and the SH angle. Figure \ref{fig:ED_comp} illustrates the SH conductivity, $\sigma_{SH}$, and SH angle, $\Theta_{SH}$, as a function of temperature for PBTTT and IDTBT. Based on our DFT calculations, the standard deviation of the distribution of the site energies for IDTBT and PBTTT are 0.64 and 0.06, respectively. The other parameters are $\alpha a=2$, $\lambda_{PBTTT/IDTBT}=0.00079$, $\nu=10^{11}$ $s^{-1}$ and $V_0=0.1$ eV. Since the two polymers are very similar otherwise, it is reasonable to assume that it is only the effect of the energetic disorder that we quantify here. It can be seen that decreasing $\epsilon$ by one order of magnitude, from IDTBT to PBTTT, increases the SH conductivity by four orders of magnitude, while the SH angle increases by a factor of 60.

The difference in the site energies appears exponential in the hopping probabilities for both electrical conductivity and SH conductivity. This explains the magnitude of the effect on the SH conductivity. We should keep in mind that a change in the energetic disorder, changes each of the conductivities to a different extent (see Eqs. \ref{econd} and \ref{shcond}). Since the SH angle is obtained via $\Theta_{sh}=\sigma_{sh}/(\sigma_{xx}\sigma_{yy})^{1/2}$, a large effect on the SH conductivity, generated by site energies, does not necessary lead to an equally large effect on the SH angle.

\begin{figure}
\includegraphics[scale=0.75]{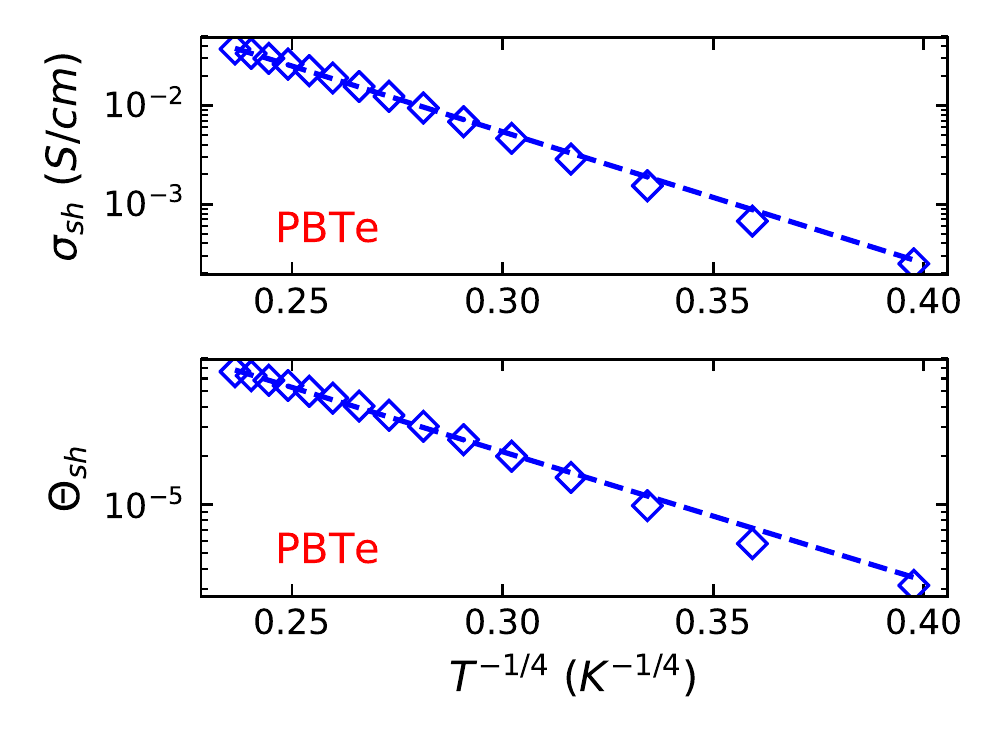}
\caption{SH conductivity $\sigma_{SH}$ and SH angle $\Theta_{SH}$ as a function of temperature for PBTe.  The temperature ranges from [40, 320] K.}
\label{fig:SOC_comp}
\end{figure}
The SOC parameter $\lambda$ that appears in the theory, is related to the spin admixture parameter, $\gamma^2$, via $\lambda=\sqrt{2\gamma^2}$ \cite{Yu2015c}. We have previously published detailed descriptions of $\gamma^2$ and its calculations from first principles \cite{yu2012spin, chopra2019accurate}.
In order to show the effect of SOC, we replace Sulphur in PBTTT  with Tellurium. The resulting polymer, PBTe, is expected to have larger SOC due to the heavy element substitution. We calculate the SOC parameter from first principles for PBTTT and PBTe,  $\lambda_{PBTTT}=0.00079$ and $\lambda_{PBTe}=0.022$. The results for the SH conductivity and the SH angle are plotted in Fig.~\ref{fig:SOC_comp}.\\
According to Fig.~\ref{fig:SOC_comp}, replacing Sulfur in PBTTT with Tellurium, PBTe, increases the SOC parameter almost 28 times and proportionally, the resulting SH conductivity and SH angle is 21 times larger. The SOC parameters does not appear in the equation for the DC electrical conductivity, Eq.~\ref{econd}, but the SH conductivity almost linearly depends on this parameter. Since the SH angle is the ratio of these two conductivities, it increases almost linearly with SOC parameter. We have previously reported the SOC parameter for almost 60 different organic molecules \cite{chopra2019chemical} for which this parameter, $\lambda$, changes from $10^{-4}$ to $10^{-1}$. This simply indicates that the SH angle could be improved by several orders of magnitude, if the right polymer is chosen.

The SH effect in organics is primarily influenced by the electron-transport and SOC, both of which intricately depend on the molecular structure. The presence of $C-C$ single bonds in these polymers allow the fragments to exist in different conformers, which also tend to increase the degree of disorder in a real polymer morphology. Larger torsion angles lead to a higher disorder that tends to hinder the charge-transport. We have recently shown that larger dihedral angles between conjugated units lead to a larger spin-admixture parameter \cite{chopra2019chemical, yu2012spin}, and therefore, larger SOC. According to our results, these two parameters combined increase the SH conductivity by more than five orders of magnitude (from IDTBT to PBTe) and the SH angle by more than three orders of magnitude (a factor of 1300).

%
%
%
%
%
%
In conclusion, we have identified important molecular characteristics with considerable impact on the SH signal in organics. These characteristics are the substitution of heavy element where a large spin density resides, and the torsion angles between constituent units of the polymer, which relate to the SOC and energetic disorder, respectively. We have extracted these parameters from first principles and we quantified the effect of each parameter on the SH effect. 
In order to obtain a larger SH signal a lower degree of energetic disorder and a larger SOC are required, both of which offer a wide range of tunability in polymeric systems. 
As it is well established, a larger SOC can be acquired via chemical substitution of a heavy element where the spin density is large. In Sulfur based chalcogenide polymers, well known for their high mobility, we predict that replacing Sulfur with Selenium or Tellurium will enhance the SH signal. The SOC can be also enhanced by increasing the torsion angles between units of the polymer. However, the latter also increases the energetic disorder which is not desirable. Based on our findings, we suggest that a group of polymers with an ordered micro-structure and substituted heavier elements, are ideal for larger SH signal. This allows for engineering easily synthesizable polymers, with much larger SH angle.%
%
%

We acknowledge financial support from the Alexander von Humboldt Foundation, the ERC Synergy Grant SC2 (no. 610115), the Deutsche Forschungsgemeinschaft (DFG, German Research Foundation)-TRR173-268565370, the Transregional Collaborative Research Center (SFB/TRR) 173 SPIN+X, and the Grant Agency of the Czech Republic grant no. 19-28375X. U.C. is a recipient of a DFG-funded position through the Excellence Initiative by the Graduate School Materials Science in Mainz (no. GSC 266).
%
%
\bibliography{References}
\end{document}